\documentclass[11pt,aps.prl,twocolumn,superscriptaddress,floatfix]{revtex4-1} 

\usepackage{amsmath}
\usepackage{graphics}
\usepackage{color}
\usepackage{caption}
\usepackage{subcaption}
\usepackage{float}
\usepackage{array}
\usepackage[margin=1.0in]{geometry}
\usepackage{ulem}
\usepackage{amssymb}

\begin{document}
\title{A Micro-Genetic Optimization Algorithm for Optimal Wavefront Shaping}
\author{Benjamin R. Anderson$^1$, Patrick Price$^1$, Ray Gunawidjaja$^1$, and Hergen Eilers$^{*}$}
\affiliation{Applied Sciences Laboratory, Institute for Shock Physics, Washington State University,
Spokane, WA 99210-1495}
\date{\today}

\email{eilers@wsu.edu}

\begin{abstract}
One of the main limitations of utilizing optimal wavefront shaping in imaging and authentication applications is the slow speed of the optimization algorithms currently being used.  To address this problem we develop a micro-genetic optimization algorithm ($\mu$GA) for optimal wavefront shaping.  We test the abilities of the $\mu$GA and make comparisons to previous algorithms (iterative and simple-genetic) by using each algorithm to optimize transmission through an opaque medium. From our experiments we find that the $\mu$GA is faster than both the iterative and simple-genetic algorithms and that both genetic algorithms are more resistant to noise and sample decoherence than the iterative algorithm.

\vspace{1em}

\end{abstract}

\maketitle

\vspace{1em}

\section{Introduction}
The ability to use wavefront shaping to control the optical properties of opaque media was first predicted by Freund in 1990 \cite{Freund90.01} and demonstrated experimentally in 2007 by Vellekoop and Mosk; who used a liquid crystal on silicon spatial light modulator (LCOS-SLM) to shape the wavefront of a laser such that the beam was focused through the material \cite{Vellekoop07.01}.   Since then the technique of using optimal wavefront shaping in conjunction with a scattering material has been used for polarization control \cite{Park12.01,Guan12.01}, spectral control of light \cite{Small12.01,Park12.02,Paudel13.01,Beijnum11.01}, enhanced fluorescence microscopy \cite{Vellekoop10.01,Wang12.01}, perfect focusing \cite{Vellekoop10.01, Putten11.01}, compression of ultrashort pulses \cite{McCabe11.01, Katz11.01}, spatio-spectral control of random lasers \cite{Leonetti13.02,Leonetti12.02,Bachelard14.01,Bachelard12.01}, and enhanced astronomical/biological imaging \cite{Mosk12.01, 
Stockbridge12.01}.  Also new techniques have been developed to aid in wavefront shaping, most notably photoacoustic wavefront shaping (PAWS) \cite{Lai14.02,Chaigne14.01,Tay14.02,Gigan14.01,Tay14.01,Lai14.01}.

One of the key features of all of these applications is the use of an optimization algorithm to search for the optimal wavefront.  The choice of which algorithm to use is crucial and determines the speed and efficiency of optimization as well as the algorithm's resistance to noise and decoherence effects \cite{Vellekoop08.01,Conkey12.02,Yilmaz13.01,Anderson14.06}. In this paper we will focus on the software (algorithm) used in optimization.  However, we note here that the hardware used also plays a major role in determining the speed of optimization, as wavefront optimization is fundamentally limited by the refresh rate of both the SLM and feedback detector. Thus far there have been three main classes of algorithms used in the literature: iterative \cite{Vellekoop07.01,Vellekoop08.01}, partitioning \cite{Vellekoop08.01}, and simple-genetic \cite{Guan12.01,Conkey12.02}.  While each method is found to have different benefits and limitations, one of the common limitations is slow optimization speed.  In order to obtain faster optimization speeds we develop a micro-genetic algorithm ($\mu$GA) for wavefront optimization as $\mu$GAs are known, from other applications \cite{Krishnakumar89.01,Senecal00.01}, to be faster than both the iterative and simple-genetic algorithms.

In this study we first describe, in detail, the operation of the iterative, simple-genetic, and micro-genetic algorithms.  We then discuss the basic differences between the algorithms and how these differences lead to the micro-genetic algorithm being the fastest algorithm.  Finally we perform an experimental comparison of the three algorithms using an optimal transmission experiment \cite{Anderson14.06} to test the algorithms' speed, efficiency, and resistance to noise and decoherence.

\section{Theory}

\subsection{Background}
Focusing light through opaque media is inherently an optimization problem in which we seek to find a specific solution (phase pattern) which optimizes the evaluation of a function (light focused in a given area).  Optimization problems have been extensively studied  and numerous techniques have been developed to approach optimization  \cite{Nocedal06.01}.  One of the simplest approaches to optimization problems is the brute force method in which every possible solution is systematically evaluated in order to find the optimal solution \cite{Nocedal06.01}.  While this technique is guaranteed to work in the absence of noise, it is extremely time consuming -- becoming unfeasible as the number of dimensions in the solution space increases -- and ineffective as the signal-to-noise ratio decreases \cite{Yilmaz13.01,Conkey12.02}.

A different approach to optimization problems, which draws inspiration from nature, are so called evolutionary algorithms (EAs) \cite{Elbeltagi05.01,Zitzler01.01,Back96.01}. EAs are stochastic algorithms in which solutions are randomly generated, evaluated, and then modified until the best solution is found.  One of the most well known classes of EAs are genetic algorithms (GAs) and they function based on the same principles as natural selection \cite{Elbeltagi05.01,Zitzler01.01,Goldberg89.01,Holland75.01,Haupt04.01}.  In GAs a population of solutions is randomly generated and then evaluated to determine how well each member solves the optimization problem. The functional response of evaluating the $i^{th}$ population member is called the member's fitness, $f_i$.  Once the members of a generation are ranked by their fitness scores, a new generation is produced through a process called crossover. During crossover the algorithm chooses two different ``parent'' solutions using a fitness based selection method 
and then the two parents are combined such that the new ``child'' solution contains half the values of one parent, and half the values of the other parent, similar to biological reproduction.  This process repeats until a new generation of solutions is produced and then the algorithm restarts with the new generation.

The standard implementation of GAs has already been applied to light transmission optimization \cite{Guan12.01,Conkey12.02} with impressive results, especially in regards to the algorithm's resistance to noise \cite{Conkey12.02}. While the standard-genetic algorithm is an improvement over the brute force iterative method, it still requires a large number of function evaluations to work, which is time consuming.  An alternative, and faster, GA implementation is the micro-genetic algorithm ($\mu$GA) \cite{Krishnakumar89.01,Senecal00.01}.  The micro-genetic algorithm primarily differs from the standard-genetic algorithm in its population size, required use of elitism (carrying over the most fit population from a generation to the next), crossover technique, and use of population resets instead of mutation.  In the following sections we will describe the SGA and $\mu$GA (as well as the iterative algorithm) in detail and expand on the differences between the two genetic algorithms.

\subsection{Iterative Algorithm}
The simplest optimization algorithm for controlling transmission is the iterative algorithm (IA) \cite{Vellekoop07.01,Vellekoop08.01}.  To begin the IA bins the SLM pixels into $N$ bins and then, starting with the first bin, the IA changes the bin's phase value in $M$ steps with a step size of $2\pi/M$ until the phase value giving the largest target intensity is found. This optimal value is then set for the bin and the procedure continues through all the bins until a phase front giving peak intensity is found.  This optimization scheme is simple to implement (requiring minimal coding) and accurately investigates the solution space. On the other hand, the algorithm requires $M\times N$ function evaluations, which quickly becomes impractical as the number of bins increases, and the IA's efficiency is severly limited by the persistence time of the sample and signal-to-noise ratio of the system \cite{Vellekoop08.01,Yilmaz13.01,Anderson14.06}.

\subsection{Simple Genetic Algorithm}
The first GA we use is a simple-genetic algorithm, which is similar to those previously used by Conkey \cite{Conkey12.02} and Guan \cite{Guan12.01}. The algorithm proceeds as follows:

\begin{enumerate}
\item Generate 30 binned random phase masks using a random number generator.
\item Evaluate the fitness of each phase mask.
\item Pass the top fifteen phase masks with highest fitness to the next generation and discard the bottom fifteen.
\item  Randomly choose two members of the fittest fifteen ($A$ and $B$ where $A \neq B$) with the probability of being chosen given by:
\begin{align}
 P_i=\frac{f_i}{\sum\limits^W_j f_j},
 \label{eqn:prob}
\end{align}
where $\sum\limits^W_j f_j$ is the net fitness of the population.  

\item Combine $A$ and $B$ into a new phase pattern $C$, where each bin in $C$ has a 50\% chance of being the value from $A$ and a 50\% chance of being from $B$.
\item During crossover, at each bin in $A$ and $B$, the two parents bin values are compared and if they are the same, a running counter is incremented. After each pair of bins has been compared this counter contains the number of bins that have the same value between the two parent images. This value is then divided by the total number of bins giving the similarity score. If the similarity scores of all parents in a generation are greater than 0.97 the bottom 15 phase masks are dropped and randomly regenerated, while the top 15 are kept.
\item  After crossover and the similarity test, the bins in $C$ are mutated at a rate of 0.005, which means that a given bin has a 0.5\% change of being assigned a new random value.
\item Steps 3-5 are repeated until the last fifteen members of the new population are created.  
\item  The fitness of the new population is measured and steps 3-6 are repeated until a stop criteria is reached (set fitness or set number of generations).
\end{enumerate}


\subsection{Microgenetic Algorithm}
The second class of genetic algorithms we consider are micro-genetic algorithms ($\mu$GAs).  $\mu$GAs have a similar structure to SGAs, but are designed to work with smaller population sizes and therefore require fewer function evaluations than SGAs.  For instance, we typically use a population of 30 phase masks for the SGA, and 5 phase masks for the $\mu$GA.   

Our $\mu$GA is structured as follows:

\begin{enumerate}
\item Generate five binned random phase masks using a random number generator.
\item Evaluate the fitness of each phase mask and rank them 1-5, with 1 having the highest fitness and 5 having the lowest fitness.
\item Discard the two phase masks with the lowest fitness scores and pass the highest-scoring phase mask to the next generation.
\item Then perform crossover on the top three ranked phase masks using the following combinations: rank 1 with rank 2, rank 1 with rank 3, and then rank 2 with rank 3 is performed twice.
\item During crossover, at each bin in the parent phase masks, the bin values of both parents are compared and if they are the same, a running counter is incremented. After each pair of bins has been compared this counter will contain the number of bins that have the exact same value between the two parent images. This value is then divided by the total number of bins giving the similarity score.  If the similarity scores of all parents in a generation are greater than 0.97 the most fit phase mask is kept and the bottom four phase masks are randomly regenerated.
\item Evaluate the fitness of the new generation.
\item Repeat steps 2 through 6. This procedure loops until a stop criteria is reached (set fitness or set number of generations)
\end{enumerate}

\subsection{Differences between the SGA and $\mu$GA.}
While the structures of the SGA and the $\mu$GA are similar -- both using randomly generated populations and crossover to explore the solution space -- there are four main differences between the two algorithms:

\begin{enumerate}
\item The $\mu$GA works with a smaller population size (five for our algorithm) than a SGA (our SGA uses thirty)
\item When choosing parents to cross over a $\mu$GA uses a tournament style selection \cite{Miller95.01,Butz02.01}, while an SGA can use a variety of different methods, such as fitness proportionate selection \cite{Back96.01}, stochastic universal sampling \cite{Baker87.01}, tournament selection \cite{Miller95.01,Butz02.01} and reward-based selection \cite{Loshchilov11.01,Deb02.01}. Our SGA uses fitness proportionate selection.
\item A $\mu$GA requires the use of elitism, while an SGA does not require it.
\item SGA's use mutation in order to avoid local maxima in the fitness landscape, while $\mu$GA's do not.  $\mu$GAs rely entirely on restarting every time the population reaches a local maxima (where the similarity score is greater than 0.97).  This is step 5 in the $\mu$GA.
\end{enumerate}

The smaller population size (and lack of mutation) allows the $\mu$GA to optimize more quickly \cite{Krishnakumar89.01,Senecal00.01} as each generation has fewer function evaluations than the SGA and the $\mu$GA skips the mutation step.  However, since the population size is smaller, the $\mu$GA requires the use of elitism to operate and continue to approach an optimal solution \cite{Krishnakumar89.01,Senecal00.01}.  Additionally, the small population size and lack of mutation leads to the $\mu$GA often drifting toward local maxima, which requires the algorithm to restart in order to escape a localized solution.

\section{Experimental Method}

\begin{figure*}
 \centering
 \begin{subfigure}[b]{0.4\textwidth}
 \includegraphics{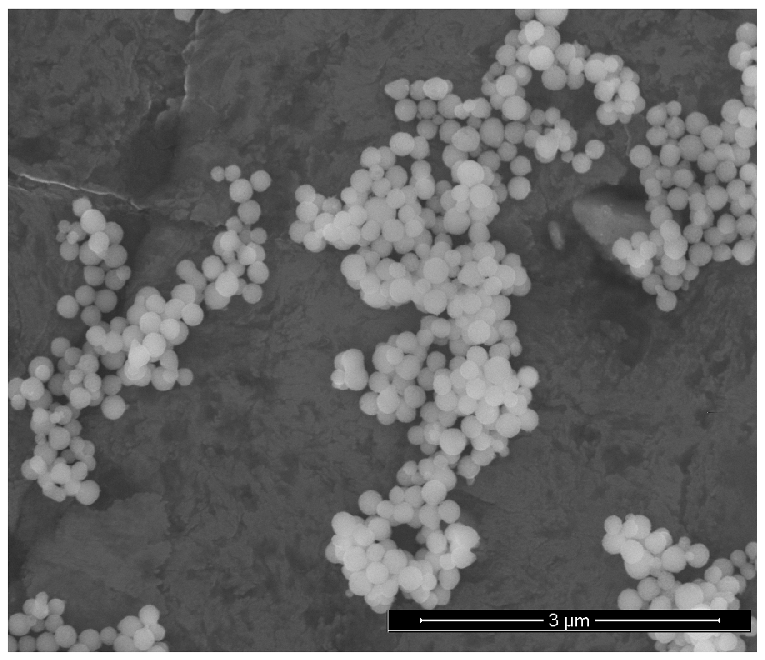}
 \end{subfigure}
 \hspace{4em}
\begin{subfigure}[b]{0.4\textwidth}
 \includegraphics{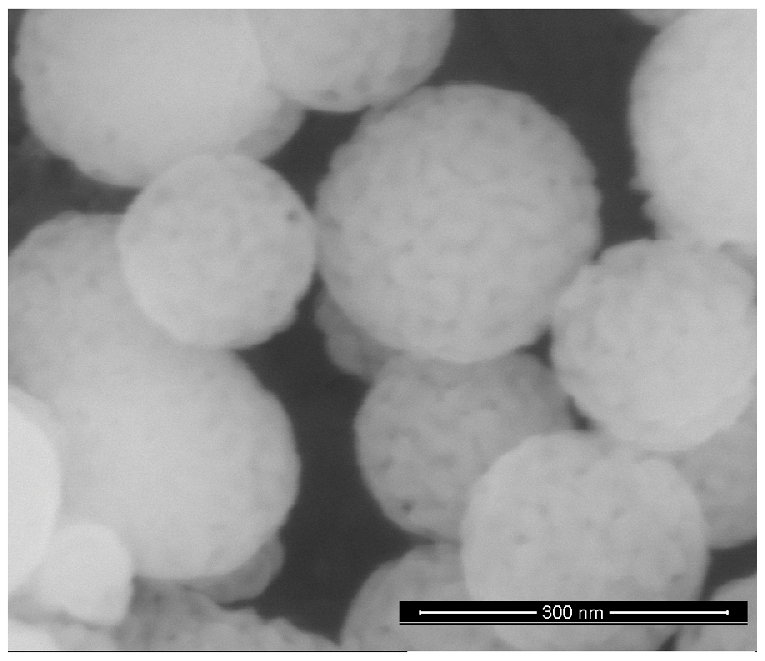}
\end{subfigure}
\caption{SEM images of ZrO$_2$ nanoparticles at magnifications of 20 000$\times$ and 200 000$\times$.  The particles are found to be spherical with diameters ranging between 80 nm and 300 nm.}
\label{fig:nps}
\end{figure*}

\begin{figure}
 \centering
 \includegraphics{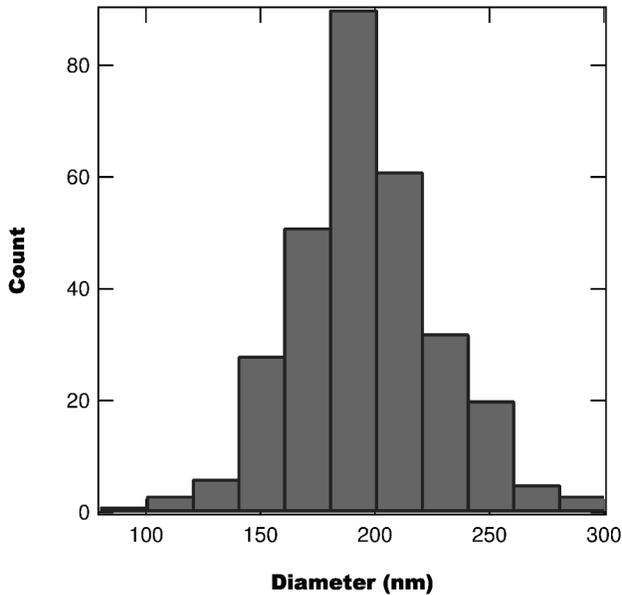}
 \caption{Histogram of NP diameters measured from SEM measurements. The mean diameter is $195 \pm 32$ nm.}
 \label{fig:sizedist}
\end{figure}

For testing the different algorithms we use two different types of opaque media: commercially available ground glass and ZrO$_2$ nanoparticle (NP) doped polyurethane nanocomposites.  To prepare the nanocomposites we first synthesize ZrO$_2$ NPs using forced hydrolysis \cite{Gunawidjaja13.01} with a calcination temperature and time of 600$^\circ$C and 1 h respectively. Figure \ref{fig:nps} shows an SEM image of the NPs at two magnifications and Figure \ref{fig:sizedist} shows the size distribution of the NPs, with the average diameter being $195 \pm 32$ nm. Once synthesized the NPs are then added to an equimolar solution of tetraethylene glycol (TEG) and poly(hexamethylene diisocyanate) (pHMDI) with a 0.1 wt\% di-n-butylin dilaurate (DBTDL) catalyst, at which point the viscous mixture is stirred and centrifuged to remove air bubbles.  The mixture is finally poured into a circular die and left to cure overnight at room temperature. 

The two sample types, ground glass and ZrO$_2$/PU nanocomposite, represent two seperate stability and scattering regiemes. The ground glass is found to have a scattering length of 970.7 $\mu$m and a persistance time on the order of days, while the nanocomposite is found to have a scattering length of 11.2 $\mu$m and a persistance time of $\approx 90$ min.  This means that the ground glass allows for more stable tests, due to its long persistance time, and larger enhancements due to its long scattering length \cite{Anderson14.06}.  On the other hand, the nanocomposite allows for tests of the algorithms' noise and decoherence resistance as the persistance time is much shorter and noise is found to have a larger effect on samples with smaller scattering lengths \cite{Anderson14.06}.

Once the samples are prepared, we use an LCOS-SLM based optimal transmission setup to compare the different algorithms.  The primary components of the setup are a Coherent Verdi V10 Nd:YVO$_4$ laser, a Boulder Nonlinear Systems high speed LCOS-SLM, and a Thorlabs CMOS camera.  See reference \cite{Anderson14.06} for more detail.  For minimal noise applications we use the ground glass sample and operate the Verdi at 10 W where the laser is most stable (optical noise $< 0.03\%$ rms) and for measuring the algorithms' resistance to noise/decoherence we use the nanocomposite sample and operate the Verdi at 200 mW where the emission is less stable (optical noise $\approx 1\%$).

\section{Results and Discussion}

\subsection{Optimization Speed}

\begin{figure}
\centering
\includegraphics{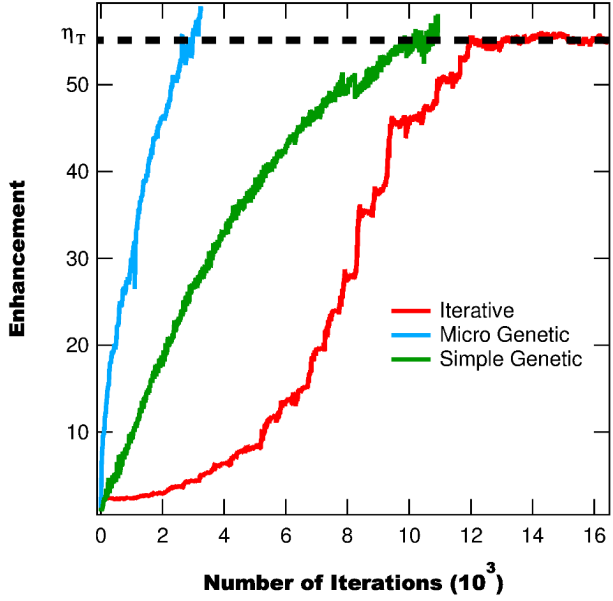}
\caption{Enhancement as a function of the number of iterations for the IA, SGA, and $\mu$GA, where the stop condition for the GAs is an enhancement of 55.  The $\mu$GA is found to be the fastest algorithm, with the SGA only being slightly faster than the IA.  }
\label{fig:AlgComp}
\end{figure}

We begin by first measuring the average intensity enhancement for the IA using bins of side length $b=16$ px, $M=32$ phase steps, and a ground glass sample.  Using the enhancement determined from the IA as the GA stop conditions, we measure the number of iterations required for each algorithm to reach the set enhancement. We use the same bin size for all three algorithms and perform five run averaging of the enhancement curves.  Figure \ref{fig:AlgComp} shows the enhancement as a function of iteration for each algorithm with the target enhancement, $\eta_T$, marked by the dashed line.  From the figure we find that both the SGA and $\mu$GA are faster than the iterative algorithm, with the $\mu$GA taking 2870 iterations, the SGA taking 10194 iterations, and the IA taking 11922 iterations.  While the SGA is found to be only slightly faster than the IA ($1.17\times$), the $\mu$GA is found to be $4.15\times$ faster than the IA, and $3.55\times$ faster than the GA for this experimental configuration.  Further 
tests with other configurations of samples and bin sizes show that the actual speed enhancement varies on the experimental parameters, but for all trials, where the three algorithms can reach the same enhancement, the $\mu$GA is the fastest algorithm.  This result is expected given the $\mu$GA's use of a smaller population size \cite{Krishnakumar89.01,Senecal00.01}.

\subsection{Resistance to Noise and Decoherence}

\begin{figure}
\centering
\includegraphics{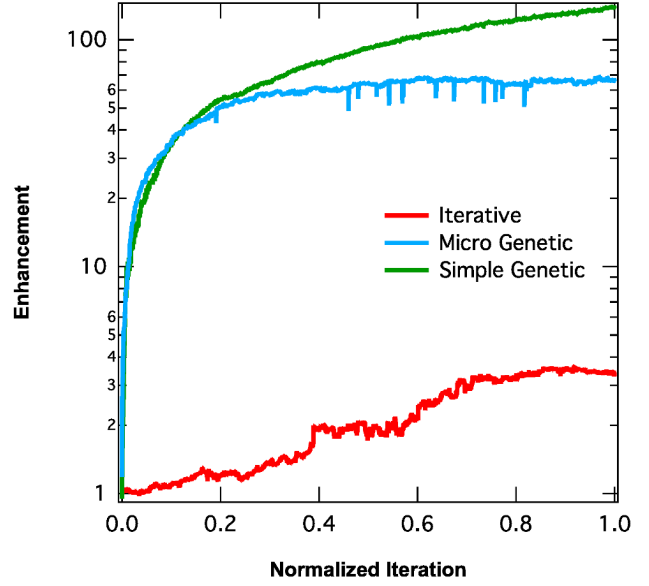}
\caption{Enhancement as a function of normalized iteration using a noisy laser, short persistence time sample, and 4096 bins.  The IA only results in a small enhancement of 3.9, while the $\mu$GA obtains an enhancement of 66 and the SGA an enhancement of 138.5.}
\label{fig:etanoise}
\end{figure}

In addition to being faster than the IA, the GAs are found to be highly resistant to the effects of both noise and sample decoherence \cite{Conkey12.02}, which are detrimental to the effectiveness of the IA \cite{Vellekoop08.01,Yilmaz13.01,Anderson14.06}.  To demonstrate the robustness of the GAs, we use a ZrO$_2$ nanoparticle-doped polyurethane sample with a short persistance time \cite{Anderson14.06} and operate the probe laser at low power where the signal-to-noise ratio is smallest. First we optimize transmission using the IA with a bin length of $b=8$ and $M=32$ phase steps, with the results being poor (shown in Figure \ref{fig:etanoise}) with a peak enhancement of 3.9.  Next we test both the SGA and $\mu$GA with bins of length $b=8$ and find drastically improved results from the IA with the $\mu$GA reaching an enhancement of 66 and the SGA reaching an enhancement of 138.5 as shown in Figure \ref{fig:etanoise} \footnote{In order to compare the 
three algorithms in Figure \ref{fig:etanoise} we plot the enhancements as a function of normalized iteration, such that the max number of iterations is set to one.  For the IA the total number of iterations is 131,072 and for both GAs the total number of iterations is 12000.}.

Figure \ref{fig:etanoise} also reveals the general functional form of the enhancement for each algorithm with the IA consisting of many sharp steps, the $\mu$GA behaving exponentially, and the SGA behaving as a power function. From Figure \ref{fig:etanoise} we find that the $\mu$GA rapidly increases and then saturates, with further iterations not adding to the enhancement, while the SGA continues to improve until the maximum number of iterations is reached. This implies that for the same number of iterations and bin size the SGA can attain larger enhancements than the $\mu$GA.

\subsection{$\mu$GA and SGA Comparison}

\begin{figure}
\centering
\includegraphics{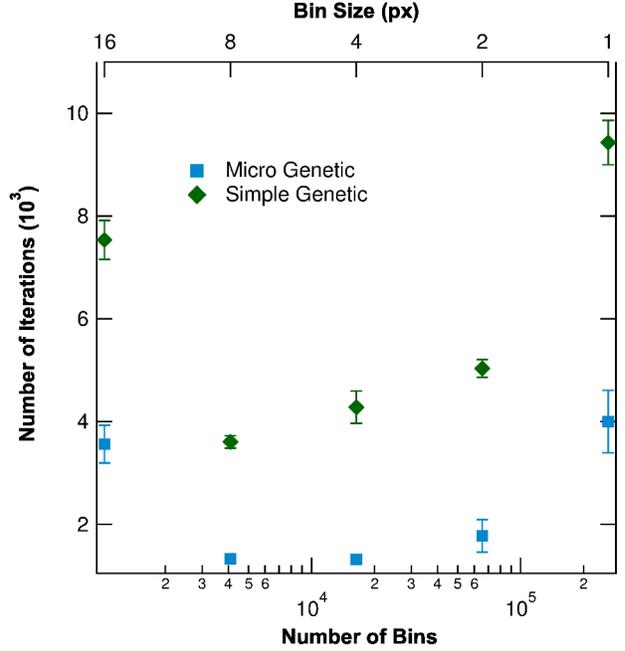}
\caption{Number of iterations required to reach an enhancement of $\eta=43$ as a function of the total number of bins for the SGA and $\mu$GA.  Bins with a side length greater than 16 px are not shown as they are found to be unable to optimize to the target enhancement within a reasonable time frame.}
\label{fig:numGA}
\end{figure}

Given the differences in the performances of the $\mu$GA and SGA in the above sections, we measure the effect of bin size on the two GA's speed and effectiveness. First we measure how bin size affects the optimization speed by using the two GAs to optimize light transmission through a polymer sample to an enhancement of $\eta=43$. A target enhancement of $\eta=43$ is chosen as both GAs can achieve that level of enhancement easily. Eventually the $\mu$GA is found to saturate, while the SGA continues to attain higher enhancements (discussed below).  This means that in order to compare the speed performace of the two GA's we need to use a target enhancment that can be reached by both algorithms.  Figure \ref{fig:numGA} shows the average number of iterations required for the two algorithms to reach the set enhancement. Only bin's of size $b\leq 16$ are shown as larger bin sizes are found to be unable to reach an enhancement of 43. For all bin sizes tested the $\mu$GA is found to be faster than 
the SGA, with a bin size of $b=8$ px resulting in the fastest optimization for both algorithms.

\begin{figure}
\centering
\includegraphics{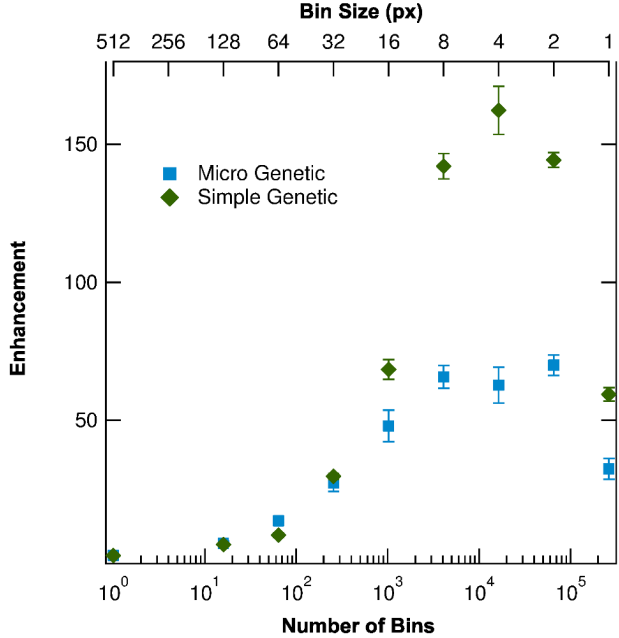}
\caption{Peak enhancement as a function of the number of SLM bins for the SGA and $\mu$GA with a stop condition of 12500 iterations.}
\label{fig:GApeak}
\end{figure}

In addition to measuring the optimization speed of the two GA's, we also consider the maximum enhancement achievable given a fixed number of iterations.  Figure \ref{fig:GApeak} shows the peak enhancement as a function of the number of bins for the GA's running 12500 iterations.  For bins of size $b\geq32$, the two GAs are found to produce the same enhancement within uncertainty.  However, as the bin size decreases the two algorithms diverge with the SGA producing greater enhancements than the $\mu$GA.  The divergence in the performance of the two GAs occurs as the SGA's larger population is better able to explore the solution domain than the $\mu$GA for small bin sizes. We also find from Figure \ref{fig:GApeak} that for the maximum number of bins ($b=1$) the enhancement is much lower than the enhancement for $b=2$.  This result is slightly counterintuitive, as having more bins should allow for a more accurate representation of the optimal phase mask.  However, by increasing the number of bins we also 
increase the solution space size, which requires more iterations to effectively explore.  Since we limit the number of iterations, we essentially limit the algorithm to searching a fraction of the total solution space, which leads to smaller enhancements.

\section{Conclusions}
We have developed a $\mu$GA for performing optimization of light through opaque media. The $\mu$GA operates based on similar principles to previously used genetic algorithms \cite{Guan12.01,Conkey12.02}, but differs in the population size, use of elitism, crossover technique, and similarity based regeneration in place of mutation. These differences lead to the $\mu$GA being significantly faster than both the SGA and IA.  This speed enhancement is advantageous for applications in which optimization must occur quickly such as biological imaging \cite{Stockbridge12.01}, authentication \cite{Anderson14.06}, and astronomical imaging \cite{Mosk12.01}. For applications where speed is less important and large enhancements are desired the SGA is the best option.

To further enhance the speed of the $\mu$GA we are currently working on implementing multi-threading into the $\mu$GA code in order to take advantage of modern computer's multi-core processors. The idea behind using multi-threading in the $\mu$GA is to perform overhead calcuations (cross over, random number generation, etc.) on one thread while a different thread controls the SLM and camera.  While this speed improvement may be small for one iteration, it will compound over the use of thousands of iterations to provide significant time savings.


\acknowledgments
This work was supported by the Defense Threat Reduction Agency, Award \# HDTRA1-13-1-0050 to Washington State University.


\end{document}